\documentclass[twocolumn]{aastex631}

\shorttitle{Flat Optically Thick Microwave Spectra}
\shortauthors{Shaik and Gary}

\usepackage{amssymb}
\usepackage{chngcntr}
\usepackage{graphicx}
\usepackage{array, makecell} 
\usepackage{amsmath}

\received{June 1, 2021}
\revised{June 25, 2021}
\accepted{June 28, 2021}
\submitjournal{ApJ}

\begin{document}
\title{Implications of Flat Optically Thick Microwave Spectra in Solar Flares for Source Size and Morphology}

\author[0000-0002-3089-3431]{Shaheda Begum Shaik}
\affiliation{Centre for Solar-Terrestrial Research \\
New Jersey Institute of Technology \\
Newark, NJ 07102, USA}

\author[0000-0003-2520-8396]{Dale E. Gary}
\affiliation{Centre for Solar-Terrestrial Research \\
New Jersey Institute of Technology \\
Newark, NJ 07102, USA}

\correspondingauthor{Shaheda Begum Shaik}
\email{ss2273@njit.edu}

\begin{abstract}
The study aims to examine the spectral dynamics of the low-frequency, optically thick gyrosynchrotron microwave emission in solar flares to determine the characteristics of the emitting source. We present the high-resolution spectra of a set of microwave bursts observed by the Expanded Owens Valley Solar Array (EOVSA) during its commissioning phase in the $2.5-18$ GHz frequency range with $1$ second time resolution. Out of the 12 events analyzed in this study, nine bursts exhibit a direct decrease with time in the optically thick spectral index $\alpha_l$, an indicator of source morphology. Particularly, five bursts display ``flat" spectrum ($\alpha_l\leq1.0$) compared to that expected for a homogeneous/uniform source ($\alpha_l\approx2.9$). These flat spectra at the low-frequencies ($<10$ GHz) can be defined as the emission from a spatially inhomogeneous source with a large area and/or with multiple emission components. In a subset of six events with partial cross-correlation data, both the events with flat spectra show a source size of $\sim120\arcsec$ at $2.6-3$ GHz. Modeling based on inhomogeneity supports the conclusion that multiple discrete sources can only reproduce a flat spectrum. We report that these flat spectra appear predominantly in the decay phase and typically grow flatter over the duration in most of the bursts, which indicates the increasing inhomogeneity and complexity of the emitting volume as the flare progresses. This large volume of flare emission filled with the trapped energetic particles is often invisible in other wavelengths, like hard X-rays, presumably due to the collisionless conditions in these regions of low ambient density and magnetic field strength.
\end{abstract}

\section{Introduction}\label{section1}
During solar flares, microwave bursts generated by gyrosynchrotron emission mechanism usually peak in 5--10~GHz frequency range \citep{Guidice1975, MelnikovDG2008} with a transition from optically thick to optically thin emission, below and above the peak, respectively. Over the decades, microwave observations conducted predominantly at optically thin frequencies have shown that the gyrosynchrotron flare emission is of mostly compact and nearly uniform sources \citep{MelnikovRSN2005,Masuda2013}. But they tend to be larger and more complicated in occasionally observed optically thick frequencies well below the peak \citep{WangDLS1994, Altyntsev2012, Gary2018}.

Solar-dedicated radio instruments like Nobeyama Radioheliograph \citep[NoRH, operating at 17 and 34~GHz;][] {Nakajima1985} have been extensively studying the high-frequency microwave (MW) flare sources. However, the low-frequency (LF) centimeter wavelength emission is less understood due to the paucity of relevant past imaging observations. There has been relatively little research on the spatial configuration and the spectral characteristics of the LF sources in the flare process. A few previous studies based on modeling and scarcely available MW imaging observations have reported large spatial sizes of the flare MW LF sources \citep{KuceraDGB1994, Fleishman2017, Fleishman2018, Kuroda2018}. Additionally, some studies have also reported large unstructured halo sources at $3$--$5$ GHz that are as large as the entire non-flaring active region  \citep{Kaltman2007, Peterova2014}.

The spectral index/slope ($\alpha_l$) of a typical gyrosynchrotron MW spectrum in the low-frequency, optically thick side (at $\nu$\textless $\nu_p$ with optical depth, $\tau$ $>$ $1$) reflects the physical parameters of the MW source region. Studies using spectral shape and slope indices have shown that the MW LF sources do not conform to the predicted spectra of a homogeneous source \citep{StahliGH1989, LeeDZ1994, MelnikovDG2008, Hwangbo2014}, and therefore are generally inhomogeneous by nature. For a spatially integrated spectrum, the source inhomogeneity results in a flatter than the expected low-frequency part of the spectrum, i.e., one with a small $\alpha_l$ value \citep{TatakuraS1970}. Studies reported that these inhomogeneous sources are mainly found to have high flux densities and large source areas at low frequencies \citep{HachenbergW1961, RamatyP1972, KleinT1984, LeeDZ1994, SongNH2016}. A few spectral studies have also shown that the evolution of simultaneous emission from multiple sources with different physical parameters can also result in a flat spectrum \citep{ShevgaonkarK1985, DulkBK1986, KleinTM1986}.

Most of the previous studies at low frequencies lack good frequency resolution, coverage, and imaging capability. There has been significantly less focus on the low-frequency part of the gyrosynchrotron emission, morphology, and behavior of the source of this emission during flares. However, the high-resolution observations in frequency and time from the Expanded Owens Valley Solar Array (EOVSA) interferometer make the data unique and valuable for spectral-based studies. The high-frequency resolution can produce finer and better-defined slopes of the spectrum compared to earlier observations. Furthermore, having corresponding imaging observations from EOVSA for one of the events is an added advantage to validate the analysis based on the spectral behavior alone, as done in the current study. 

In this paper, we first address the spectral characteristics observed in a set of 12 bursts during the peak of the solar cycle 24 (in 2015). We focus on the occurrence of the flat spectra at low frequencies, and we determine the source area by analyzing the observed flux density spectra. For the frequency range of $\sim2.5$ to $18$ GHz, we use the calibrated total power and uncalibrated cross-correlation data during EOVSA's commissioning phase. At this early time, due to the absence of the auto-correlation data transferring the total power calibration to the cross-correlated visibility data is prevented. Instead, we analyze the source size information inherent in the visibility data by forming a relative measure (a pseudo-relative-visibility described in \S\ref{sec:RV}). In addition, to demonstrate the role of inhomogeneity on the flat spectrum, we implement inhomogeneous modeling to generate the observed spectrum.

\section{Data and Methods} \label{section2}
The primary data used in this study are the pre-imaging spectral measurements from the commissioning phase of EOVSA, located near Big Pine, CA \citep{GaryEOVSA2014, Gary2018}. EOVSA comprises thirteen 2.1-m antennas with a frequency resolution of about 40~MHz and a temporal resolution of 1~s in the MW frequency range of 2.5--18~GHz. During this commissioning phase, EOVSA observed a number of flares in total power mode only, from January to June 2015, and then added a partial cross-correlation mode (data from a limited number of baselines). EOVSA had eight antennas and was running two independent copies of a prototype 4-element correlator design. During this time, the prototype correlator recorded only 12 baselines and did not produce the correct auto-correlation data. EOVSA attained full imaging capability with a 16-element correlator starting in April 2017. We include one event with imaging data (2017 September 10) to validate our methods for interpreting the earlier, less-complete data.

The total power calibration is performed based on the daily flux density measurements reported by the National Oceanic and Atmospheric Administration (NOAA) from the U.S. Air Force Radio Solar Telescope Network (RSTN) and Penticton at nine frequencies (eight from RSTN and one from Penticton). The calibration procedure is to read these daily flux density measurements, calculate the mean value at each frequency, fit a quadratic function to the fluxes at seven frequencies in the $1-15.4$ GHz range, and apply interpolation or extrapolation to match the EOVSA frequencies. Any pointing offsets are determined for all the antennas as a function of frequency and are used with primary beam corrections to the observed data.

Radio imaging data available from the Nobeyama Radioheliograph (NoRH) at 17 GHz are complemented with the EOVSA data for one of the events (2015 March 10 M2.9 flare) shown in this paper. NoRH has the capability of imaging $17$ and $34$ GHz optically thin emission with a spatial resolution of about $10$ and $5\arcsec$ respectively \citep{Nakajima1985, Shibasaki1994}. The hard X-ray emission, which is believed to correlate with the MW emission from the same or closely related electron population, is obtained from the Reuven Ramaty High Energy Solar Spectroscopic Imager (RHESSI; \citealt{Lin2002}) for spatial comparison. In addition, for determining the configuration of the photospheric and coronal magnetic fields in the flaring region, the extreme ultraviolet (EUV) images from the Atmospheric Imaging Assembly (AIA; \citealt{Lemen2012}) and magnetograms from the Helioseismic and Magnetic Imager (HMI; \citealt{Scherrer2012}) onboard the Solar Dynamics Observatory (SDO) are correlated.

The set of 12 bursts analyzed in this study are listed in Table~\ref{table1}. Most bursts have a short burst duration ranging from as low as 1 minute to 6~minutes and smooth time evolution of the flux density spectra. For a burst observation, each antenna measures the same total power spectrum independently. The median over the operating antennas is utilized to arrive at a single dynamic spectrum of the burst. The standard deviation among antennas is used as a measure of instrumental uncertainty. 

Each burst in the list is processed for flagging bad antennas, background subtraction, flux calibration, and corrections for temporal discrepancies in the data for a few bursts. Then the spectral fitting is performed on the observed spectra for the whole duration of each burst. This fitting provides a reliable set of parameters for a large number of time points at each frequency from the given high time and frequency resolution of EOVSA. The parameters are obtained from the procedure as described and introduced in \cite{StahliGH1989}. The functional form of this procedure for the flux density is written as 
\begin{equation}
S({\nu}) = A \nu^{a}(1-e^{-B\nu^{-b}}).
\end{equation}
This equation for the generic shape of the gyrosynchrotron MW spectrum provides a positive slope at low frequencies (optical depth, $\tau$ $>$ $1$), reaches a peak flux density ($\nu_p$, peak frequency in the range $5$ to $10$ GHz) and forms a negative slope at high frequencies ($\tau$ $<$ $1$). At low frequency, the term $e^{-B\nu^{-b}}$ becomes negligible for coefficients $B$ and $b$ making $(1-e^{-B\nu^{-b}}) \approx 1$. Therefore, the low-frequency slope ($\alpha_l$) of the MW spectrum is well represented by the parameter $a$. Similarly, the other parameters deduced from the equation are high-frequency slope ($a-b$), peak flux $S(\nu_{p}$), and peak frequency $\nu_{p}$. This generic functional fitting is carried out on each 1~s of observed spectrum for the events in Table~\ref{table1}, thus providing the temporal evolution of all the parameters. 
Furthermore, the spatial components of the MW sources are determined from the technique of relative visibility \citep{GaryH1989,KuceraDGB1994,Kuroda2018}. The relative visibility (henceforth referred to as RV) can be deduced from the observed visibility data measured by an interferometer, even when a lack of phase calibration prevents true imaging. RV is the normalized Fourier transform of the observed visibility amplitudes. Mathematically, RV can be derived from the ratio of cross- and auto-correlations $\frac{x_{ij}}{\sqrt{a_{ii}a_{jj}}}$, where $x_{ij}$ and $a_{ii}$, $a_{jj}$ are the cross- and auto-correlated amplitudes respectively, for $i$ and $j$ antennas. Alternatively, it can be written as the ratio of fringe amplitude $v(s)$ at antenna spacing $s$ (in wavelengths) to the total power amplitude $v(0)$, zero spacing. 

RV can give an estimate of the one-dimensional source size of the burst under the assumption of a Gaussian source brightness distribution. For this Gaussian source with flux density $S(x)= pe^{-(x-x_0)^2/\alpha^2}$, where $p$ is peak flux at angular position $x = x_0$, the logarithm of the relative visibility is 
\begin{equation} \label{eq:2}
\begin{aligned}
\ln(RV) = \ln \bigg(\frac{v\tiny(s\tiny)}{v\tiny(0\tiny)} \bigg) = -9.325 \times 10^{-14} d^2 B_{\rm cm}^2 f_{\rm GHz}^2 \\
= -8.393 \times 10^{-11} B_\lambda^2 d^2,
\end{aligned}
\end{equation}
\noindent where the visibility $v(s) = \int_{-\infty}^{\infty}p e^{-(x-x_0)^2/\alpha^2}e^{-2\pi isx} dx = p \sqrt\pi\alpha e^{-\pi^2s^2\alpha^2}e^{-2\pi ix_0s}$. Therefore, a plot of logarithm of the RV as a function of square of projected baseline length (distance between each pair of antennas) produces a line whose slope is directly proportional to the size of the source at every frequency. This source size $d$ is the one-dimensional full width at half maximum (FWHM), $B_\lambda$ and $B_{\rm cm}$ are the baseline lengths in wavelength and cm units, respectively, $f_{\rm GHz}$ is the observing frequency in GHz, and $s$ is the spatial frequency or inverse of the fringe spacing. The FWHM size $d$ is related with the Gaussian 1/$e$ width $\alpha$ as $d = 2\sqrt{\ln2}\, \alpha = 1.665\,\alpha$.
 
Then, the slope $m$ of the $\ln{(RV)}$ versus $B_\lambda^2$ plot derived from Equation~(\ref{eq:2}), is given by 
\begin{equation} \label{eq:3}
m(\nu) = -8.393 \times 10^{-11} \times d(\nu)^2,
\end{equation}
which gives the source size in arcsec as

\begin{equation}  \label{eq:4}
d = \sqrt{m(\nu) (-1.192 \times 10^{10})} \, .
\end{equation}

\noindent Thus, from the observed visibility of each burst, the relative visibility expressed as a function of frequency can estimate a Gaussian-equivalent source size at each frequency as discussed in  \S\ref{sec:RV}.

\begin{deluxetable*}{ccccccc}
\tablenum{1}
\tablecaption{List of selected bursts observed by EOVSA\label{table1}}
\tablewidth{0pt}
\tablehead{
\colhead{Event $\#$} & \colhead{Date} & \colhead{GOES class} & \colhead{Start time} &
\colhead{NOAA active region} & \colhead{Region configuration} & \colhead{Peak flux}\\
\colhead{} & \colhead{} & \colhead{} & \colhead{(UT)} & \colhead{} & \colhead{} & \colhead{ (sfu)}
}
\startdata
$1^\ast$  & Mar 10, 2015 & M2.9 & 23:46     & 12297       & $\beta\gamma\delta$ & 1338 \\
$2^\ast$  &Mar 12, 2015  & M2.7 & 21:44     & 12297       & $\beta\gamma\delta$ & 548 \\
3         &Apr 21, 2015  & M2.0 & 16:55     & 12322       & $\beta$             & 97\\
4         &Apr 21, 2015  & M1.8 & 21:39     & 12322       & $\beta$             & 31\\
$5^\ast$  &May 5, 2015   & X2.7 & 22:05     & 12339       & $\beta\gamma$       & 1441\\
$6^{\ast\ast}$&Jun 21, 2015 & M2.0 & 01:02  & 12371       & $\beta\gamma\delta$ & 1252\\
7         &Aug 22, 2015  & M3.5 & 21:19     & 12403       & $\beta\gamma$       & 985\\
$8^\ast$  &Aug 24, 2015  & M1.0 & 17:40     & 12403       & $\beta\gamma\delta$ & 256\\
9         &Aug 24, 2015  & C3.0 & 22:40     & 12403       & $\beta\gamma\delta$ & 253\\
10        &Sep 24, 2015  & C3.3 & 23:41     & 12418       & $\alpha$            & 76\\
11        &Sep 27, 2015  & C4.2 & 17:41     & 12422       & $\beta\gamma\delta$ & 388\\
$12^\ast$ &Sep 27, 2015  & C9.3 & 19:44     & 12422       & $\beta\gamma\delta$ & 126
\enddata
\tablecomments{Asterisks mark the flat spectral events and the double asterisk marks event 6 with flat spectrum only in the peak phase. (sfu, solar flux unit is a measure of solar radio flux density; 1 sfu = $10^4$ {\rm Jy} = $10^{-22}$ W\,m$^{-2}$Hz$^{-1}$)}
\end{deluxetable*}

\section{Observations}
\subsection{Spectral Index and Flat Spectra} \label{sec:flat spectrum}
The low-frequency optically thick spectral index $\alpha_l$, as discussed in Section~\ref{section1}, is a sensitive parameter of the MW burst spectrum that indicates the source spatial characteristics as a function of frequency. This slope and the shape of the non-thermal gyrosynchrotron radiation spectrum are conveniently defined by numerical approximations to the theory \citep{DulkM1982, Dulk1985} for an isolated homogeneous source. The relationship between the brightness temperature $T_b$, the effective temperature of the radiating electrons $T_{eff}$, and the emitted flux density $S$ of the radio source are given in the following equations for the optically thick regime:
\begin{equation} \label{eq:5}
T_b = T_{\rm eff}, 
\end{equation}
\begin{equation} \label{eq:6}
T_{\rm eff} = 2.2 \times 10^{9-0.31\delta} (\sin\theta)^{-0.36-0.06\delta}\\ \bigg(\frac{\nu}{\nu_B} 
\bigg)^{0.5+0.085\delta},
\end{equation}
where $\nu$ and $\nu_B$ are the observed frequency and gyrofrequency respectively, $\theta$ is the viewing angle and $\delta$ is the electron spectral index.
In addition, the total power flux density is the brightness temperature integrated over the source,
\begin{equation} \label{eq:7}
S({\nu}) =  \frac{2k \nu^{2}}{c^2} \int T_b({\nu})\,d\Omega  \quad  [Wm^{-2}Hz^{-1}],
\end{equation}
where $k$ and $c$ are the Boltzmann constant and velocity of light, $d\Omega$ is the differential solid angle in steradians and $T_b$ is the effective temperature in Kelvin for the optically-thick emission as in equation~(\ref{eq:5}).

For the typically observed values of $\delta$ in the range of $2 \leq \delta \leq 7$ and from Equations~(\ref{eq:5}) to (\ref{eq:7}), \[S({\nu}) \propto \nu^{2+x}\] with $0.75 \leq x \leq 1.095$. 

\begin{figure*}
\centering
\includegraphics[width=2\columnwidth]{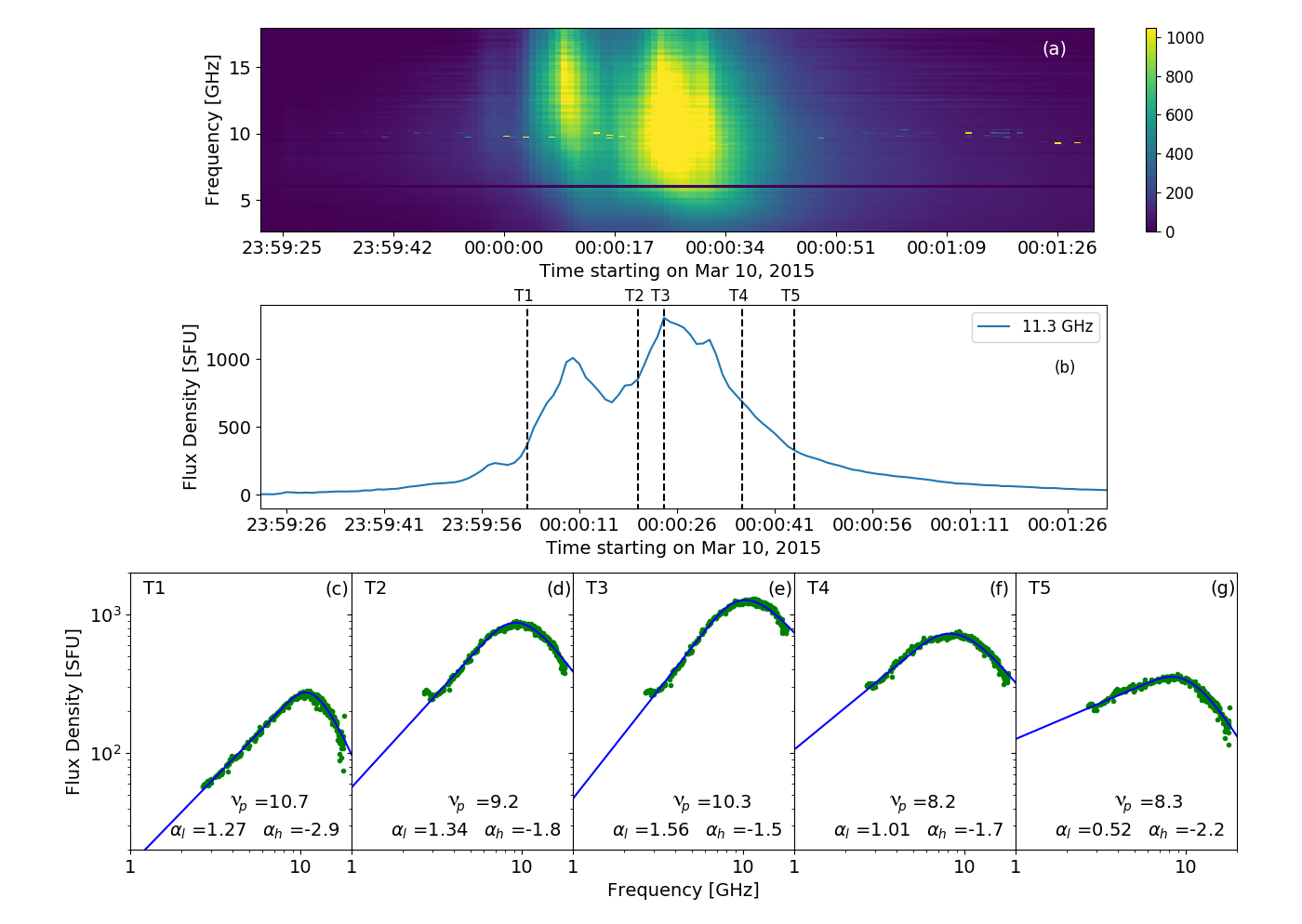}
\caption{MW spectral evolution of the 2015 March 10 event. (a) Total power median dynamic spectrum for the frequency range 2.5--18~GHz. (b) Time profile at 11.3~GHz (near the burst peak frequency). (c to g) Spectra for the five times, T1, T2, T3, T4, and T5 marked in (b) with vertical dashed lines. T1 and T2 are chosen for the rise phase of the pre-peak (00:00:09 UT) and the main peak (00:00:24 UT), respectively. Spectral fit parameters are noted at the bottom of each panel.}\label{fig:fig1}
\end{figure*}
Therefore, for a single homogeneous burst source, one would expect the observed flux density spectrum to have an optically thick slope $\alpha_l \approx 2.75$--3.1 for $\delta \approx 2$--7. The average $\alpha_l$ of this relatively small range can thus be taken as typical value for a homogeneous source,
\begin{equation} \label{eq:8}
S({\nu}) \propto \nu^{2.9}.
\end{equation}
Any value of the slope far from this range indicates some peculiarity in the characteristics of the source. Steeper spectra, $\alpha_l \geq 3.1$ can only be due to Razin suppression or absorption by a different source of cooler, intervening plasma \citep{Ramaty1969,Bastian2007}. Shallow/flat spectra with $\alpha_l \leq 2.75$ indicate spatial inhomogeneity of the source emission. Therefore, examining the low-frequency spectral index of the gyrosynchrotron spectrum provides a sensitive means to reveal complexity in the source morphology. 

It should be noted that $\alpha_l\approx2.9$ is for a non-thermal distribution of electrons. Emission from a hot ($T\sim10$ MK) thermal distribution can produce a low-frequency slope of 2 but, for a homogeneous source, it would also produce an extremely steep high-frequency slope \citep{GaryH1989}. Since our bursts do not show such steep high-frequency slopes and have low-frequency slopes shallower than 2, we interpret the emission as non-thermal and inhomogeneous.

Initially, all the events in Table~\ref{table1} are analyzed for the low-frequency index $\alpha_l$ of their spectra. The procedure is illustrated in Figure~\ref{fig:fig1}, which gives the overview of the first event, 2015 March 10 in Table~\ref{table1}. Figure~\ref{fig:fig1}a shows the total power dynamic spectrum over a 2 minute period, while Figure~\ref{fig:fig1}b shows the flux density time profile at the peak frequency, 11.3~GHz. The corresponding flux density spectral evolution is shown in Figure~\ref{fig:fig1}c--g. The vertical dashed lines in Figure~\ref{fig:fig1}b indicate five times in the burst---2 times during the rise phase, the peak time, and two times in the decay phase---selected for the spectra shown in Figure~\ref{fig:fig1}c--g and designated as T1 to T5. For each time, the spectral fitting is applied using the St{\"a}hli equation as discussed in Section~\ref{section2}. The fitting parameters, low-frequency index $\alpha_l$, high-frequency index $\alpha_h$, and peak frequency $\nu_p$ are marked at the bottom of each spectral plot. 

In Figure~\ref{fig:fig1}c--g, the main point of interest of this study, the low-frequency spectral index always remains well below the theoretical value of around 2.9 predicted for a homogeneous source by Equation~(\ref{eq:8}). This low index value is observed more pronounced in the decay phase of the burst. Additionally, in agreement with the previous studies \citep{LeeDZ1994, MelnikovDG2008}, the peak frequency $\nu_p$ clearly increases in the rise phase (T2 to T3) and decreases during the decay phase, which is an indicator \citep{MelnikovDG2008} that the peak is controlled by optically thick gyrosynchrotron emission. 
\begin{figure*}	
\includegraphics[width = 2\columnwidth]{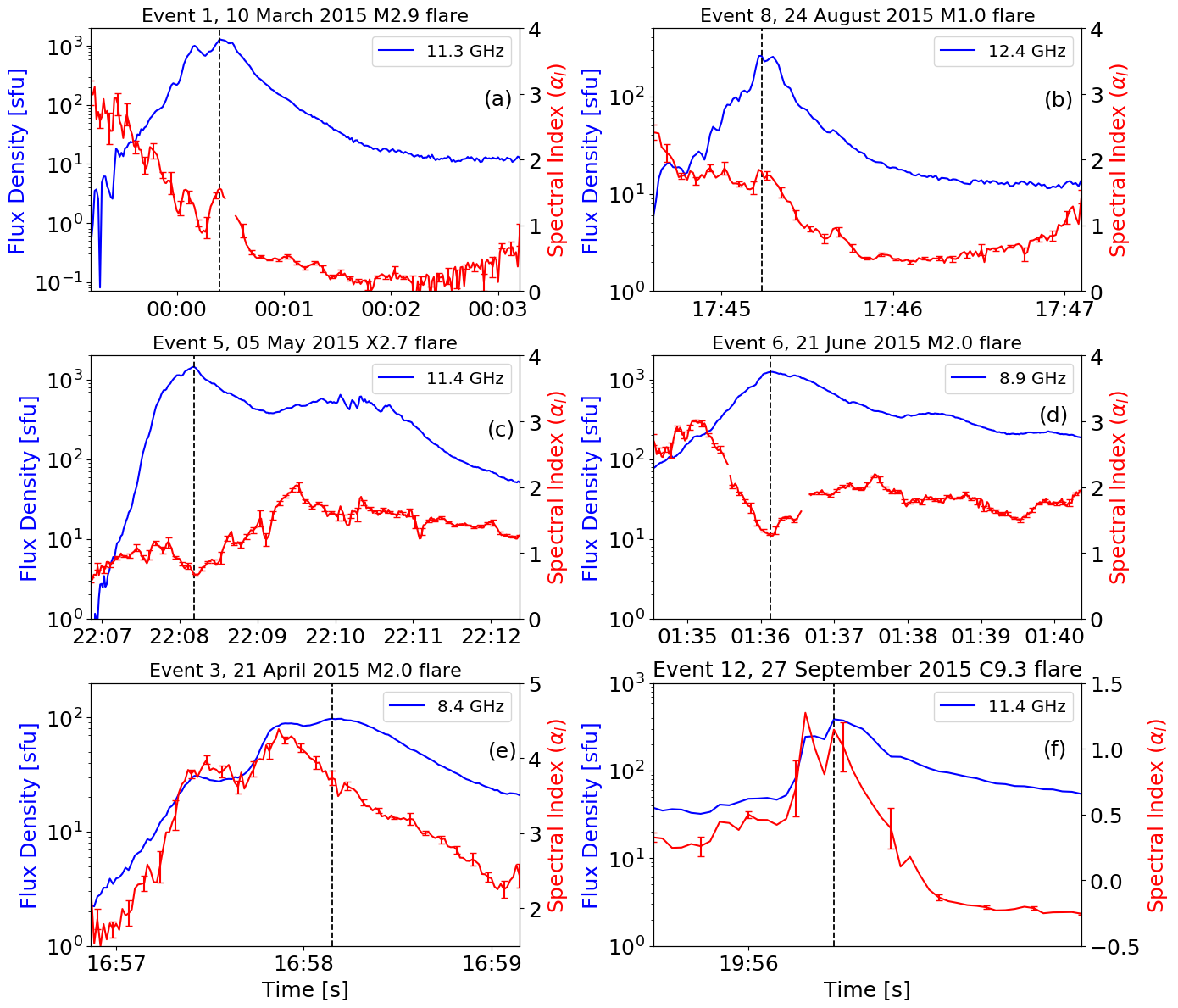}
\caption{Spectral index evolution of the six representative events for the whole duration of the bursts (red). The corresponding flux density time profile are plotted in blue color. Note that the time profiles are plotted in logarithmic scale in y-axis.}	\label{fig:fig2}
\end{figure*}

To statistically examine $\alpha_l$ and its evolution over the duration of the bursts, the time profiles of $\alpha_l$ for all 12 events are determined in the same manner. Figure~\ref{fig:fig2} shows $\alpha_l$ over the duration of six of the bursts selected for their representative trends. In each panel, the spectral index is plotted in red (scale on the right side of each plot) and the flux density at the peak frequency in blue (scale on the left). The error bars in the spectral index curves are calculated from the standard deviation over the neighboring 5~s (5 data points).

\begin{figure*}
\includegraphics[width=2\columnwidth]{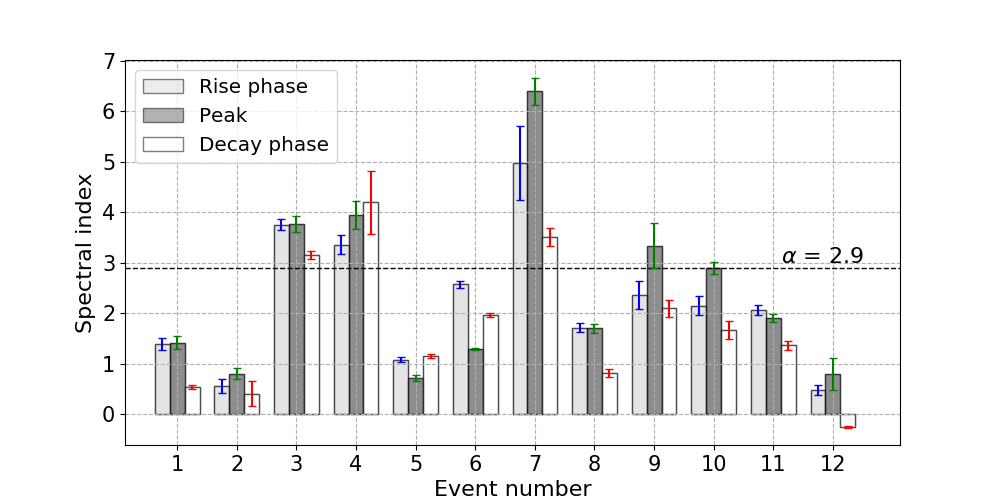}
\caption{Bar chart of spectral index for the events in Table~\ref{table1}. In each event, the spectral indices are shown at each of the rise, peak and decay phases (see text for details). The homogeneous source model spectral index of 2.9 is marked as a black dashed horizontal line.}
\label{fig:fig3}
\end{figure*}

The overall behavior of the plots illustrates three types of $\alpha_l$ evolutionary trend. In the first type, as shown in Figure~\ref{fig:fig2}a and b, $\alpha_l$ starts to decrease before the peak and continues to drop into the decay phase except for a slight, short-duration increase near the peak. During the late decay phase, $\alpha_l$ shifts back to higher values. For example, in the 2015 March 10 event shown in Figure~\ref{fig:fig2}a the index starts with a value of $\sim 3$ in the rise phase then smoothly reduces to a value of $1$ at peak and eventually decreases almost to $0$ in the decay phase. In the second type (Figure~\ref{fig:fig2}c and d), $\alpha_l$ shows a sharp decrease at the peak and increases in the early decay phase. In the third type of $\alpha_l$ evolution, shown in Figure~\ref{fig:fig2}e and f, the spectral index follows the same trend as the flux density, growing steeper during the rise phase and dropping rapidly after the main peak and in the decay phase.

To better compare the behavior of the spectral index over all the events, Figure~\ref{fig:fig3} presents a bar chart for the trend of $\alpha_l$ versus each event number listed in Table~\ref{table1}. The chart reports the indices averaged over a 10 s period in the rise phase, peak, and decay phase of each event (e.g., at times T1, T3, and T5 in Figure~\ref{fig:fig1} corresponding to each phase). The rise phase for a given event is defined as the time when the flux density first reaches 10\% of the maximum, while the decay phase is at the same flux density during the decay. The theoretical homogeneous source spectral index of $\sim$2.9 is marked by the horizontal dashed line. Note that the $\alpha_l$ value in each phase of the burst differs slightly from the value shown in Figure~\ref{fig:fig1} for the 2015 March 10 event due to the averaging of the ten values in each corresponding phase in Figure~\ref{fig:fig3}. The error bars are the standard deviation of fluctuations in $\alpha_l$ during each 10 s period.

The observations of the chart are summarized as follows.
\begin{enumerate}
\item \textit{Index evolution:} Over the evolution of the burst, the spectral index decreases from peak to decay phase for nine out of the 12 events (Event numbers 1--3, 7--12).

\item \textit{Index value:} Compared to the theoretical value, most of the events have $\alpha_l$ below 2.9. Nine of the events have spectral indices less than 2.9 in at least one of the three phases or all the phases of the burst (Event numbers 1, 2, 5, 6, 8--12). In particular, five of them (1, 2, 5, 8, and 12) show extreme spectral index values $<$ 1 in at least one phase of the burst, which we henceforth define as flat events.

\item Contrary to that suggested in earlier studies, flat low-frequency spectral indices are not limited to large, high-flux-density (X-class) flares \citep{HachenbergW1961,RamatyP1972, LeeDZ1994} but can also appear in a relatively weak C9.3 flare (event 12) with a low peak flux $\sim 125$ sfu.

\item All the events but one (event 5, the sole X-class flare in our list) that show flat indices originate from active regions with a complex magnetic configuration of $\beta\gamma\delta$ as seen in Table~\ref{table1}.

\item The variation of the spectral index within an event is generally smaller than between the events. An event with either a large or small index maintains similar values during its evolution.
\end{enumerate}

Interpreting these points in terms of the homogeneity of the emitting source, a very low value of $\alpha_l$ signifies that the emission is from a highly inhomogeneous source. The area/emitting volume of this source grows with decreasing frequency due to the non-uniform physical parameters within. The declining value of $\alpha_l$ with time in some events suggests that the inhomogeneity and the complexity of the burst source grow as the flare evolves with time.

This spectral index analysis leads to a representative set of 5 events (42$\%$ of our sample) that have a flat ($\alpha_l<1$) spectrum, which is further investigated for additional evidence of source inhomogeneity in the next section.

\section{Results and Discussion}
\subsection{Source Area Spectrum} \label{section4.1}
The MW flux density as a function of frequency $S({\nu})$ for a simple homogeneous source as mentioned in Equation~(\ref{eq:7}) can be written as
\begin{equation} \label{eq:9}
S({\nu}) =  \frac{2k \nu^{2} T_b({\nu})}{c^2} \Omega(\nu)\quad [Wm^{-2}Hz^{-1}].
\end{equation}
At the observed frequency $\nu$, for a constant brightness temperature $T_b$, the emitted flux density $S({\nu})$ is directly proportional to the solid angle area of the source $\Omega(\nu)$ ($\Omega$ as a function of $\nu$ emphasizes that the source area indeed depends and changes with frequency). We expect that most of the flares observed in MW emission exhibit some level of inhomogeneity, leading to an increase in source area with decreasing frequency. The flat events that we have identified require an extreme rise in size and hence inhomogeneity.

Equation (\ref{eq:9}) can be rewritten following Equation (4) in \cite{Fleishman2016b, Fleishman2018} as
\begin{equation} \label{eq:10}
    A\approx137\,\frac{S(\nu)[{\rm sfu}]}{\nu^2_{\rm GHz}}\Bigl(\frac{10^8\,{\rm K}}{T_b}\Bigr),
\end{equation}
\noindent where $A$ is the area in square arcsec, the constant applies when the flux density is expressed in sfu, and the frequency is in GHz. Note that the constant factor is corrected and is different from the equation in \cite{Fleishman2016b, Fleishman2018} due to an over-simplification in their expression (Fleishman, private communication). As discussed earlier, the non-thermal brightness temperature for the optically thick part of the spectrum is equal to the effective temperature, $T_b = T_{\rm eff}$. Observations show that the effective temperature during large flares is typically quite high; thus, the last term is of order unity. If we assume a fixed, frequency-independent value for the effective temperature ($\gg 10^7$ K), we can obtain an approximate representation of the source area spectrum for the optically thick regime of the observed flux density. For a given spectrum, according to Equation~(\ref{eq:8}), $S \propto \nu^{\alpha_l}_{\rm GHz}$. Therefore, Equation~(\ref{eq:10}) leads to the source area $A \propto \nu^{\alpha_l-2}_{\rm GHz}$. 

\begin{figure*}
\centering
\includegraphics[width=2\columnwidth]{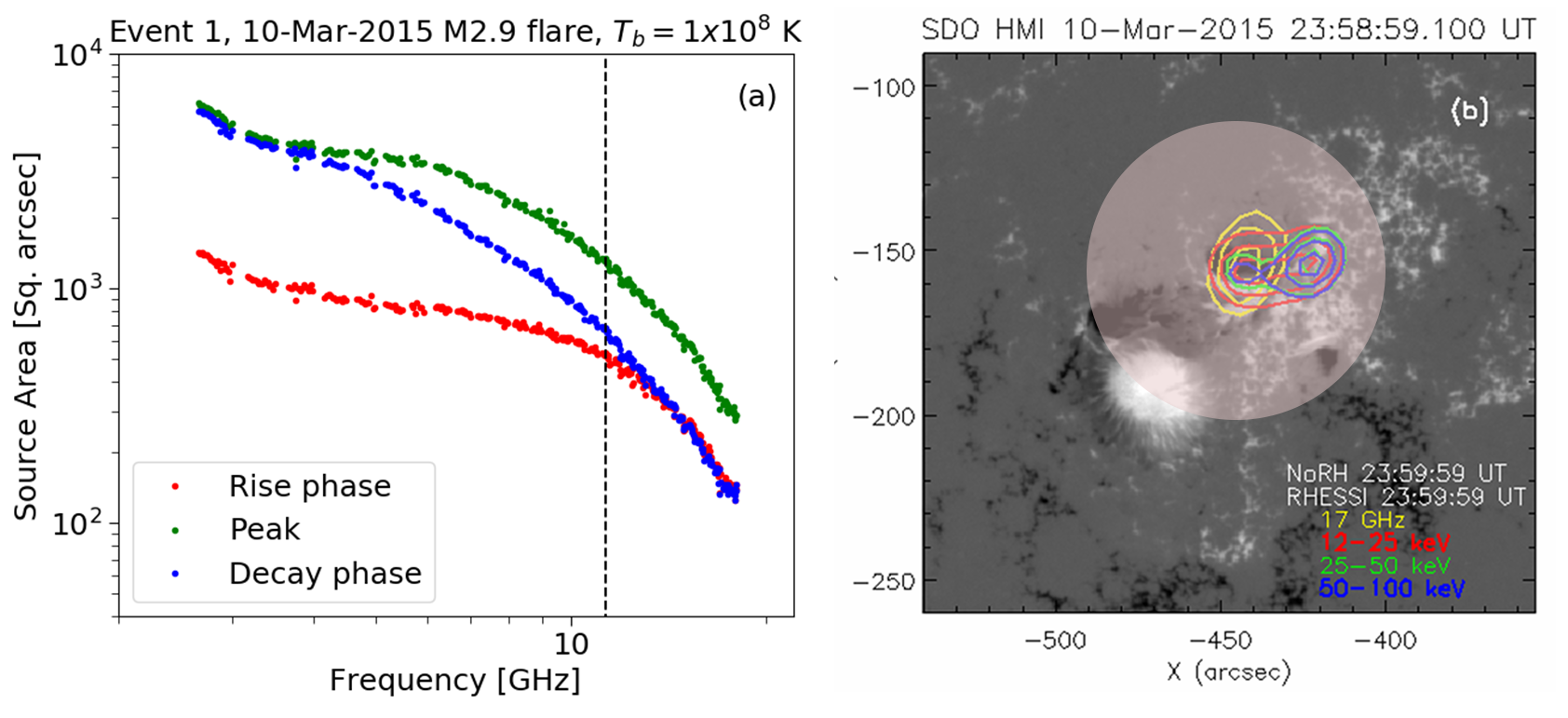}
\caption {Source area spectrum at each phase of the burst and the image map of the flare sources at the peak time. (a) The source area calculated for a given brightness temperature is marked in red, green, and blue for the three phases. The peak frequency, $11.3$ GHz, is marked by the vertical dashed line to separate optically thick and thin parts of the spectrum. (b) NoRH and RHESSI flare emission at $50$, $70$, $90\%$ of their corresponding maximum fluxes are plotted over the HMI magnetogram at the burst peak time. The masked region shows the equivalent circular MW source area cartoon for 2.9 GHz.}
\label{fig:fig4}
\end{figure*}

As an illustration, Figure~\ref{fig:fig4}a shows the source area spectrum of the 2015 March 10 burst, whose $\alpha_l$ varies from 1.3 to 0.5 over the burst duration. For the observed EOVSA flux density, we assume a constant and high brightness temperature of $10^8$ K and calculate the source areas as shown in the figure. The three curves are the areas measured for the times selected in Figure~\ref{fig:fig1} (T1, T3, and T5) at the rise phase, peak, and decay phase, respectively. These source size estimates are valid for the optically thick emission well below the spectral peak, shown by the vertical dashed line. In the peak (green) and decay phase (blue) curves, at the low frequencies $\sim2.9$ GHz, the source areas are $\sim6200$ and $\sim5500$ arcsec$^2$, respectively. The area spectrum for both the peak and decay phase is relatively flat until $\sim4$ GHz and then decreases more steeply in the decay phase from frequencies above $\sim4$ GHz. For the rise phase (red) curve, the source area starts smaller than the other two phases, at $\sim1500$ arcsec$^2$. However, this size would be larger if we make the reasonable assumption of a variable effective temperature, starting from a lower value and increasing with time towards the peak of the burst. Following the source area at one frequency, say $3$ GHz, the rise phase area starts with a large value and grows still larger in the peak and decay phase. 

According to Equation~(\ref{eq:10}), for the case of a flat spectrum with $\alpha_l\approx 0.5$, the source area goes as $A\propto \nu^{-1.5}_{\rm GHz}$. For example, in the decay phase, a moderate-sized source with area $\sim 900$ arcsec$^2$ at $\nu_p\sim10$~GHz must grow to an area of $\sim 5000$ arcsec$^2$ at 3~GHz. For the range of spectral index values $\alpha_l$ observed in this event, the power-law index $n$ in $A\propto \nu^{-n}_{\rm GHz}$ ranges from $0.7$ to $1.5$.

At $2.9$ GHz, the area for the peak time gives a diameter of $\sim89\arcsec$ assuming a circular shape of the source (light gray circle in Figure~\ref{fig:fig4}b). For comparison, the high-frequency NoRH MW 17~GHz images and the RHESSI Hard X-ray (12--25, 25--50, and 50--100 KeV) sources at the burst peak time are overlaid on a HMI magnetogram as shown in the figure. So, even with our assumption of a high brightness temperature, the estimate of the low-frequency radio source size is many times greater than the 17~GHz and the hard X-ray sources (which themselves may appear larger than they are due to finite resolution). The actual brightness temperature in the flaring site, if not as high as the assumed $T_b$, will only lead to a much larger source.

As the emission is optically thick over its volume, these source area measurements characterize the actual area of the source magnetic structure. The changes observed in the area spectrum can be caused by gradients in the magnetic field strength and density \citep{KleinT1984,KuceraDGB1994,Bastian1999,Fleishman2018} that result in spatially-dependent changes in opacity. The fact that the flux density spectrum becoming flatter with time and the source area spectrum becoming steeper indicates that the sources grow significantly large with decreasing frequencies. Such large sources cannot be homogeneous but have to be non-uniform and inhomogeneous in the flare site. This line of reasoning with inhomogeneity is further discussed in Section~\ref{section4.3}. 

We have performed a similar analysis on all the other events and verified that the flat events exhibit the same trend of large areas. We now seek confirmation of these estimates by indirect interferometric measurements via the relative visibility technique described earlier. Before doing that, however, in the next section, we examine the RV technique by comparing its results with the direct EOVSA imaging of source sizes available for the 2017 September 10 event \citep{Gary2018}.

\subsection{EOVSA Relative Visibility Analysis}\label{section4.2}
As discussed earlier, relative visibility is a sensitive measure of source size and complexity for a flare MW emission using the observed visibility amplitudes. For a Gaussian source, RV amplitude vs. baseline length shows a smoothly decreasing shape depending only on the source size \citep{GaryH1989}. For an extended source, RV is unity at short baselines and decreases with increasing baseline length. Generally, short baselines cannot resolve a single Gaussian source; therefore, cross and auto-correlated data will have almost the same flux density leading their ratio to be unity. When sources become resolved at longer baselines, the cross-correlated data has less power leading to the RV ratio gradually decreasing from unity. Any deviation of the source from a Gaussian shape will modify the manner in which the RV ratio decreases, but the initial drop at short baselines is expected to measure the size of an equivalent Gaussian shape.
\subsubsection{2017 September 10, X8.2 class Flare}
\begin{figure*}
\includegraphics[width=2\columnwidth]{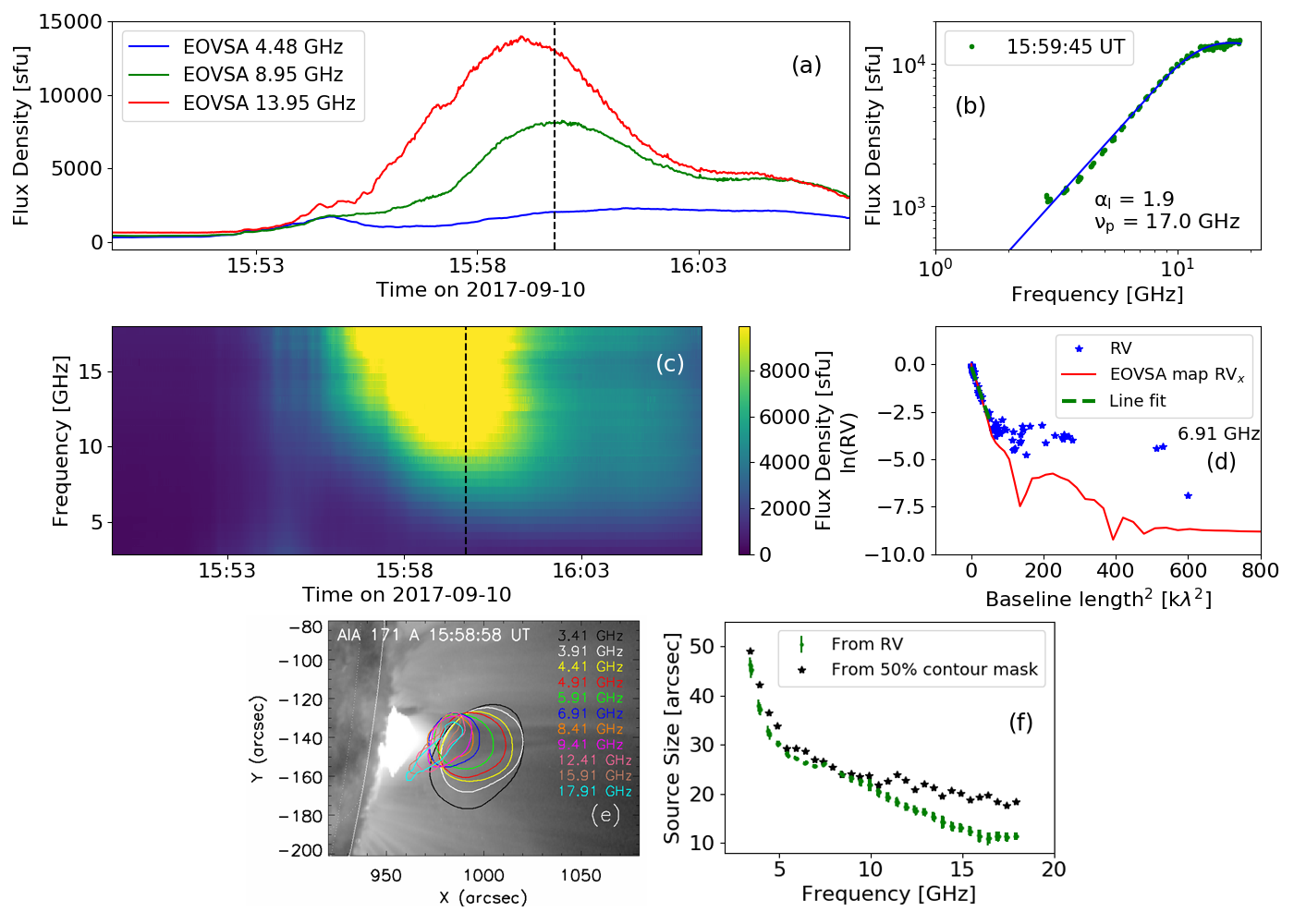}
\caption {Relative visibility and source size analysis at 15:59:05 UT around the peak of 2017 September 10 X8.2 flare at the three given frequencies. (a) to (c): Time profiles, flux density spectrum at the peak time (with $\alpha_l$ and $\nu_p$ marked at the bottom of the plot), and median total power dynamic spectrum of the burst. (d) $\ln(RV)$ versus $B_\lambda^2$ plot at the peak time for 6.91 GHz with a linear fitting marked in green. RV calculated solely from the observed EOVSA image maps is in red. (e) The $50\%$ contours of peak flux density of EOVSA images at the given frequencies are overlaid on AIA $171$ {\AA} EUV map. The solar limb is marked in white. (f) Source size spectrum deduced from the relative visibility slopes are plotted in green and the images in black. The error bars on the green symbols are the uncertainties of the fitting coefficients calculated from the covariance matrix from the fitting procedure.}
\label{fig:fig5}
\end{figure*}

The 2017 September 10, X8.2 class west limb flare is one of the largest flares in solar cycle 24 that occurred in the active region AR 12673 \citep{Gary2018, Omodei2018}. The time profiles of the MW burst at three selected frequencies and the total power dynamic spectrum are shown in Figure~\ref{fig:fig5}a and~\ref{fig:fig5}c, respectively. This event is a long duration burst extending for more than an hour with a gradually evolving rise phase and a long decay phase. The black dashed vertical line in Figure~\ref{fig:fig5}a marks the time of the 8.95~GHz peak used for the relative visibility analysis. Figure~\ref{fig:fig5}b shows the flux density spectrum for this time with the fitting procedure of the curve as in Figure~\ref{fig:fig1}. The spectral index $\alpha_l$ and peak frequency $\nu_p$ are marked at the bottom of the panel. Although not important for the RV validity check, we remark that this event is intermediate between a homogeneous ($\alpha_l =2.9$) and a flat spectral type ($\alpha_l < 1$) with $\alpha_l$ changing from $1.9$ to $1.4$ over the time shown in Figure~\ref{fig:fig5}a. At the chosen time, the spectrum shows broadband emission with the spectrum still rising at 18~GHz, which implies that the peak frequency would have occurred beyond 18 GHz.

Using the cross- and auto-correlated data relation for RV, the logarithm of RV (blue dots) versus square of baseline length ($B_\lambda^2$) at the selected time is shown in Figure~\ref{fig:fig5}d. As noted above, the approximately linear decrease in $\ln(RV)$ at short baselines is the behavior expected for a Gaussian source, whose fitted slope (green dashed line) provides an estimate of source size as per Equation~\ref{eq:3}.  This procedure can be repeated at each of the 30 frequency bands available in the data to create a source size spectrum. However, Figure~\ref{fig:fig5}d shows that after following the line for several e-foldings, the points begin to deviate from a single slope at longer baselines. To examine this non-linear distribution of RV amplitudes at longer baselines, we obtain the alternative RV (red curve) directly from the available EOVSA images for this event shown in Figure~\ref{fig:fig5}e. The contours are for a subset of available frequencies at the selected time overlaid on the bright AIA $171$ {\AA} EUV loops.

The image RV (red curve) is obtained from the visibility space by taking the Fast Fourier Transform (FFT) of the image map. The RV is calculated for the row of pixels along the FFT plane's horizontal axis (referred to as EOVSA map $RV_x$), with the zeroth element used to supply the zero spacing intensity needed in Equation~\ref{eq:2}. When plotted to the square of baseline lengths at 6.91 GHz, these relative visibilities resemble the true RV, and both show a dip near  120 k$\lambda^2$ (with a small dip $\sim$ 35 k$\lambda^2$) and a maximum near 200 k$\lambda^2$. We interpret this as evidence for a tendency of the source to have a more uniform surface brightness and sharper edges than a true Gaussian source so that its FFT develops sinc-function-like lobes. When examined at other frequencies, the overall pattern persists for each increasing frequency. This pattern shifts in a regular manner towards the longer baseline lengths, increasing the width of each lobe in the sinc function and decreasing the slope value of the linear region. Both of these changes are indicative of the source size growing smaller with increasing frequency, consistent with the images in Figure~\ref{fig:fig5}e.

Finally, to deduce the quantitative source size measurements from the actual RV, the linear portion of the RV distribution in Figure~\ref{fig:fig5}d is passed through a linear fitting procedure. The green dashed line shows the fit at 6.91~GHz after restricting the fit to the inner 40 baselines (out of total 78 baselines). These are the baselines that sample the linear portion of RV plots for the full frequency range. The FWHM source size is then determined from the slopes of these fits using Equation~(\ref{eq:4}) to generate a source size spectrum as shown in Figure~\ref{fig:fig5}f in green symbols.
Additionally, the one-dimensional circular size measured directly from the $50\%$ contour of the peak intensity of the images is overplotted in black symbols. Both the measures agree reasonably well up to 10~GHz, after which the source sizes diverge mainly could be due to our assumption that the source is circular, whereas the EOVSA emission as seen in Figure~\ref{fig:fig5}e becomes more elliptical and elongated. 

This exercise demonstrates that the RV source spectrum is a reliable tool to deduce the source size as a function of frequency even in the absence of imaging spectroscopy, so long as the source approximates a Gaussian shape.  Furthermore, the departures from a Gaussian profile are easily recognized from the RV data. Having validated the RV approach, we now apply the RV analysis to the events in Table~\ref{table1} and examine them for the source morphology differences between flat events and the non-flat normal events.

\subsubsection{RV Analysis of Flares in our Sample Set}\label{sec:RV}
As discussed earlier, the auto-correlation measurements are not available for the set of bursts in our study, which were taken with a prototype correlator that was not producing correct auto-correlations. Therefore, we must form a pseudo-RV by substituting the data from one of the short inner baselines in the place of the auto-correlation data. Generally, the pseudo-RV is determined by $\frac{x_{ij}}{x_{\rm short}}$, where $x_{\rm short}$ is the cross-correlated amplitude of any sufficiently short baseline. This short baseline has a frequency-dependent fringe spacing ($>$18--2.5~arcmin for 2.5--18~GHz) large enough to guarantee that any reasonable flare source is unresolved. We note that, unfortunately, one of the strengths of RV--that it is independent of calibration--is lost for this pseudo-RV form on those events, so we must limit our study to the events with good gain calibration. Due to that, events 1-6 in Table~\ref{table1} are not ideal for this pseudo-RV analysis and therefore limiting to events 7-13.

\begin{figure*}
\includegraphics[width=2\columnwidth]{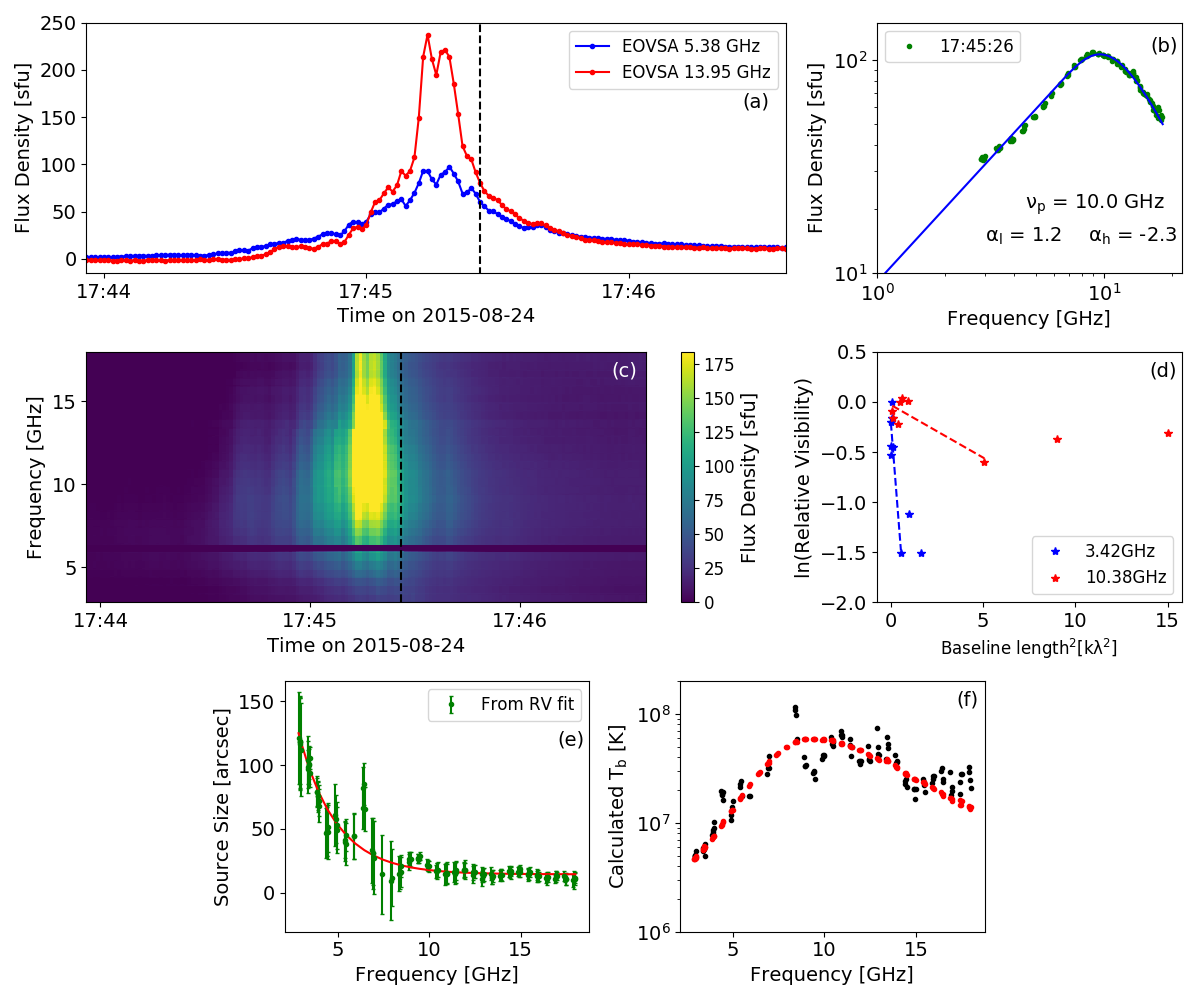}
\caption{Relative visibility and source size analysis during the decay phase of the August 24, 2015 flare. (a) to (c): Time profiles, flux density spectrum at the decay phase, and the total power median dynamic spectrum of the burst. Fitting parameters are marked at the bottom of the panel (b). (d) ln(RV) as a function of $B_\lambda^2$ at the two given frequencies. (e) Source size spectrum extracted from relative visibility slopes at each frequency. The red curve is the exponential form fitting of the source size. (f) Calculated brightness temperature spectrum from the obtained source size measurements. The black symbols are from the actual size measurements (green in panel (e)), and the red symbols are from the fitted curve (red in panel (e)).}
\label{fig:fig6}
\end{figure*}

To illustrate the steps in the pseudo-RV analysis, we use the decay phase of the M1.0 class flare observed on 2015 August 24 (a flat spectral event, number 8 in Table~\ref{table1}, with an averaged index of $\sim$1.8 at the peak and $\sim$ 0.8 in the decay phase). Figure~\ref{fig:fig6}a and c show the time profiles at 7.88 and 13.95~GHz, and the total power dynamic spectrum of the burst, respectively. The time profiles indicate that the event is a very impulsive burst lasting for about a minute. The vertical dashed line marks the time 17:45:26 UT, at which the pseudo-RV is determined, selected in the decay phase with flux density $>25\%$ of the peak value. Figure~\ref{fig:fig6}b is the flux density spectrum at this time, with the low frequency slope $\alpha_l = 1.2$.

The number of good baselines available for this event was 9, far fewer than the 2017 September 10 event with 78 baselines. Figure~\ref{fig:fig6}d is the $\ln(RV)$ vs. square of the baseline length plot at the two frequencies marked at the bottom of the panel. Here, the pseudo-RV is calculated using $\frac{x_{ij}}{x_{14}}$, where ${x_{14}}$ is the cross-correlated data from antennas 1 and 4 acting as an auto-correlation component which has the maximum peak flux density compared to the other short baselines. For both the frequencies plotted in the figure, the 9 RV amplitude points are spread such that the first 6 are in a close cluster at short baselines, and the remaining 3 are spread outwards at the longer baselines.

When these $\ln(RV)$ vs. $B_\lambda^2$ plots are viewed progressively with frequency, the 9 RV points shift in a fashion similar to that of the previous event, with the points extending outward for increasing baseline length. Even though a clear sinc-like function curve cannot be distinguished due to the smaller number of baselines, the same combination of linear and non-linear trends is apparent. The fitting is carried out with the first 7 points, which lie on the linear trend, ignoring the flatter trend of points 8 and 9, which may represent the sinc-like function in Figure~\ref{fig:fig5}. Fits to these 7 points for the given two frequencies are shown by the dashed lines in Figure~\ref{fig:fig6}d. Fits at other frequencies follow this same trend, giving us confidence that these fits reveal the general source size trend with frequency despite the relatively large scatter of the points. 

By determining the slopes of the line fits at each frequency, the source size spectrum is obtained as shown in Figure~\ref{fig:fig6}e, where the large error bars reflect the uncertainties in the individual fits. The source size starts with a value of $\sim125\arcsec$ at the lowest frequency $2.9$ GHz and continues to decrease to $\sim10$ GHz. After $\sim10$ GHz, the size remains small and almost constant for higher frequencies, as expected for an optically thin source. The overall pattern of source size with frequency is well fitted with an exponential function (red curve in Figure~\ref{fig:fig6}e).

\begin{figure*}
\centering
\includegraphics[width = 2\columnwidth]{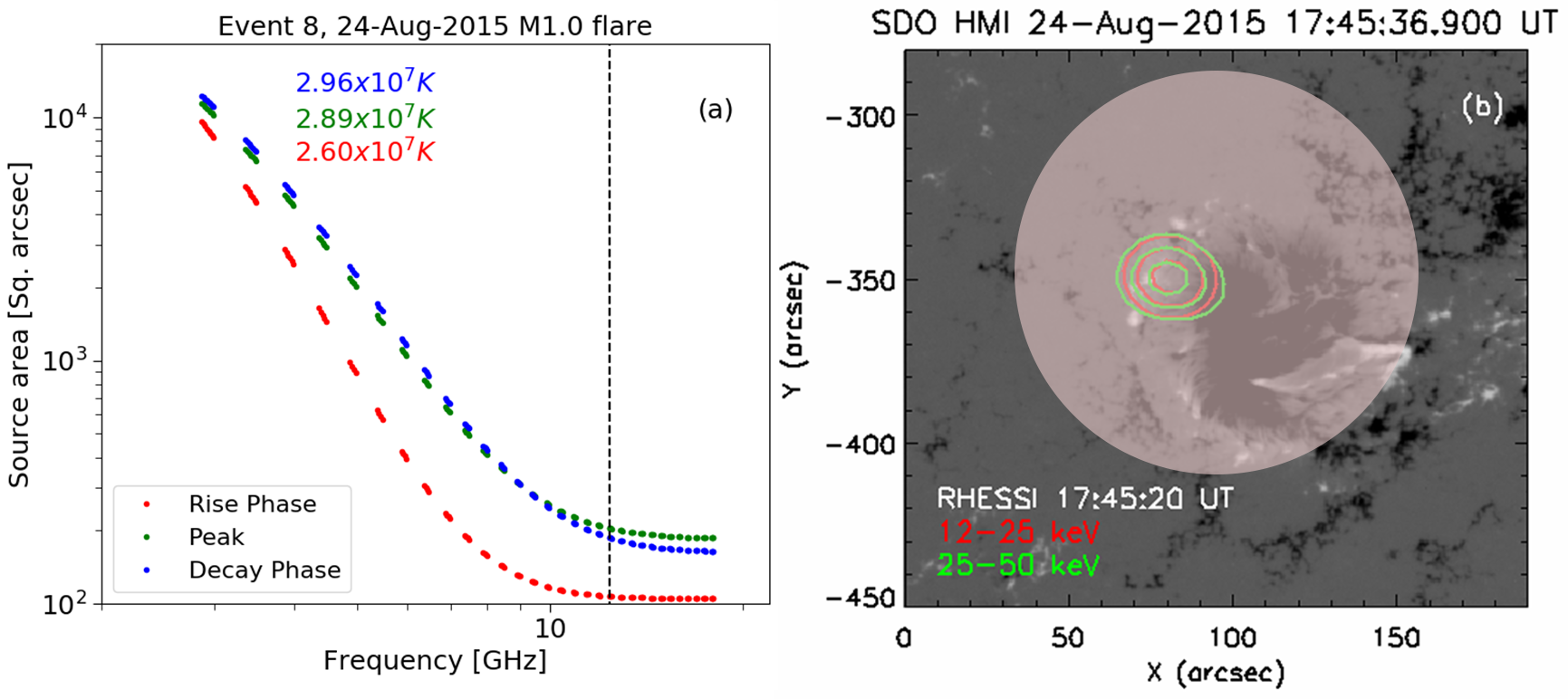}
\caption {Source area spectrum from RV measurements at the three phases of the 2015 August 24 burst with the corresponding hard X-ray emission. (a) The areas are obtained from the RV applied at each of the three phases (red, green, and blue). The brightness temperatures marked at each phase are the average values calculated from the RV measurements. The peak frequency, $12.4$ GHz at the peak time, is marked by the vertical dashed line. (b) RHESSI flare emission contours are plotted at 50, 70, 90$\%$ of peak flux in red and green over the HMI magnetogram. The decay phase LF source at $2.9$ GHz is shown as a cartoon with a circular masked region.}
\label{fig:fig7}
\end{figure*}
The source size estimates from the RV analysis can be used to calculate the brightness temperature spectrum for the measured flux density as shown in Figure~\ref{fig:fig6}f. The peak brightness temperature reaches at least $6\times10^7$~K and decreases towards both lower and higher frequencies. The black points in Figure~\ref{fig:fig6}f are calculated from the individual points in Figure~\ref{fig:fig6}e, while the red points are calculated from the exponential fit.  

Figure~\ref{fig:fig6} gives the RV source measurements for one time during the decay phase of the burst. We repeat the analysis for the other phases of the burst (rise and peak) and convert the exponentially fit source size to arrive at the source area spectra assuming a circular source shown in Figure~\ref{fig:fig7}a (red, green, and blue points, respectively). The measurements suggest that the source was already quite large during the rise phase and grew substantially larger at the peak. Then the source stopped evolving in size and faded in brightness during the decay phase.

It is of interest to compare the source area spectrum in Figure~\ref{fig:fig7}a with the similar one in Figure~\ref{fig:fig4}a for the 2015 March 10 event. Recall that the source area spectrum in Figure~\ref{fig:fig4}a was derived from an assumption of a single constant brightness temperature over the whole frequency range. This assumption served to show that the source area must be large, but it cannot be realistic and leads to a sudden flattening of the curves at lower frequencies in Figure~\ref{fig:fig4}a instead of the continued rise in the source area, we see in Figure~\ref{fig:fig7}a. With the benefit of RV analysis, we could derive a brightness temperature spectrum that varies with frequency in agreement with expectations from theory \citep{Dulk1985} as in Figure~\ref{fig:fig6}f. For this decreasing $T_b$ at the low frequencies, the source area continues to rise steeply and shows a much larger source area needed to match the observed flux density.

For spatial comparison, the one-dimensional size of $\sim 125\arcsec$ at $2.9$ GHz from the RV measurements for the decay phase is overlaid as a cartoon on an HMI magnetogram in Figure~\ref{fig:fig7}b, with RHESSI contours at $12-25$ and $25-50$ keV. Clearly, the low-frequency emission of the flare requires a far larger source extent than the RHESSI contours ($\sim 30\arcsec$). Most of the usually reported RHESSI flare sources, as seen in Figure~\ref{fig:fig4}b, are restricted to only the higher density regions that occur at low coronal heights in the flare \citep{Kosugi1991,Benz2008, Krucker2010}. The extended emission that we observe here suggests the involvement of the overlying magnetic structures, which have correspondingly lower magnetic field strength and density, leading to such low-frequency emission. The conclusion that LF sources sometimes exhibit a large emission area at a relatively high brightness temperature $>10^7$ K agrees well with the previous recent studies (e.g., \citealt{Fleishman2017, Fleishman2018}).

This pseudo-RV analysis is further conducted on the remaining events listed in Table~\ref{table1} (events 7 and 9 to 12) to estimate their source morphology. The results are shown in Table~\ref{table2} along with the spectral index at the time of RV measurements. These source sizes shown here are those measured for the lowest frequency observed in each of the events. The flat spectral events, in particular, have shown a source size of $\ge$ $120\arcsec$, and thus, there is generally an anti-correlation between low-frequency source size and spectral index.
\begin{deluxetable}{ccc}
\tablenum{2}
\tablecaption{Source size measurements from RV averaged over 3 seconds at the lowest frequency\label{table2}}
\tablewidth{0pt}
\tablehead{
\colhead{Event $\#$} & \colhead{Spectral index $\alpha_l$} & \colhead{Source size}\\
\colhead{} & \colhead{} & \colhead{(arcsec)}
}
\startdata
$1^\ast$&1.5 & - \\
$2^\ast$&0.8 & - \\
3 & 3.7 & - \\
4 & 3.7 & - \\
$5^\ast$ & 0.7 & - \\
$6^{\ast\ast}$ & 1.3 & - \\
7 & 6.6 & 70 \\
$8^\ast$ & 1.2 & 125 \\
9 & 2.1 & 97 \\
10 & 3.0 & 71 \\
11 & 1.8 & 95 \\
$12^\ast$ & 1.1 & 120
\enddata
\tablecomments{The corresponding averaged spectral index of each event at the time of RV calculation is given in the second column.}
\end{deluxetable}
\subsection{Adding Inhomogeneity}\label{section4.3}
We have demonstrated that emission from a large source area at low frequencies is needed to make the flux density spectrum flat. We have suggested that this is due to a rather extreme source inhomogeneity. As a flat MW spectrum diverges from the spectrum produced by a single uniform source, the general homogeneous source theoretical model cannot produce an acceptable fit. Hence, modeling that includes inhomogeneity of the source is needed to explain the observed flat spectrum \citep{KleinTM1986}. An intermediate step in complexity is to consider an inhomogeneous model consisting of multiple homogeneous sources that are physically discrete, but when combined, will result in the observed flat spectrum. Thus to demonstrate the idea of inhomogeneity, we adopt a model introduced by \cite{Hwangbo2014}, which represents the three-dimensional (3D) magnetic field distribution with multiple sources of homogeneous components.  

The model uses seven components with different magnetic field strengths and physical parameters. These components are simultaneously adjusted to generate a composite spectrum that matches the observed spectrum. The primary factors defined here for the modeling are the source area, $A$ (perpendicular to the line of sight or LOS) and the thickness, $L$ (along the LOS) of each component using two ad hoc scaling laws: area of each segment
\begin{equation} \label{eq:11}
A_i = A_0\bigg({\frac{B_i}{B_0}}\bigg)^{-\alpha},
\end{equation}
\noindent and thickness of each segment
\begin{equation}\label{eq:12}
L_i = L_0\bigg(\frac{B_i}{B_0}\bigg)^{-\beta} \text {with i = 0 to 6.}
\end{equation}

The magnetic field strength is logarithmically scaled between $B_0$ and $B_6$, with $B_6$ being the component with the smallest magnetic field strength. The gyrosynchrotron emission for the components of Equations~\ref{eq:11} and~\ref{eq:12} is derived from Dulk's approximations \citep{Dulk1985} using Equations $13$ to $17$ of \citealt{DulkM1982}.

\begin{figure}[h]
\includegraphics[width=1.1\columnwidth]{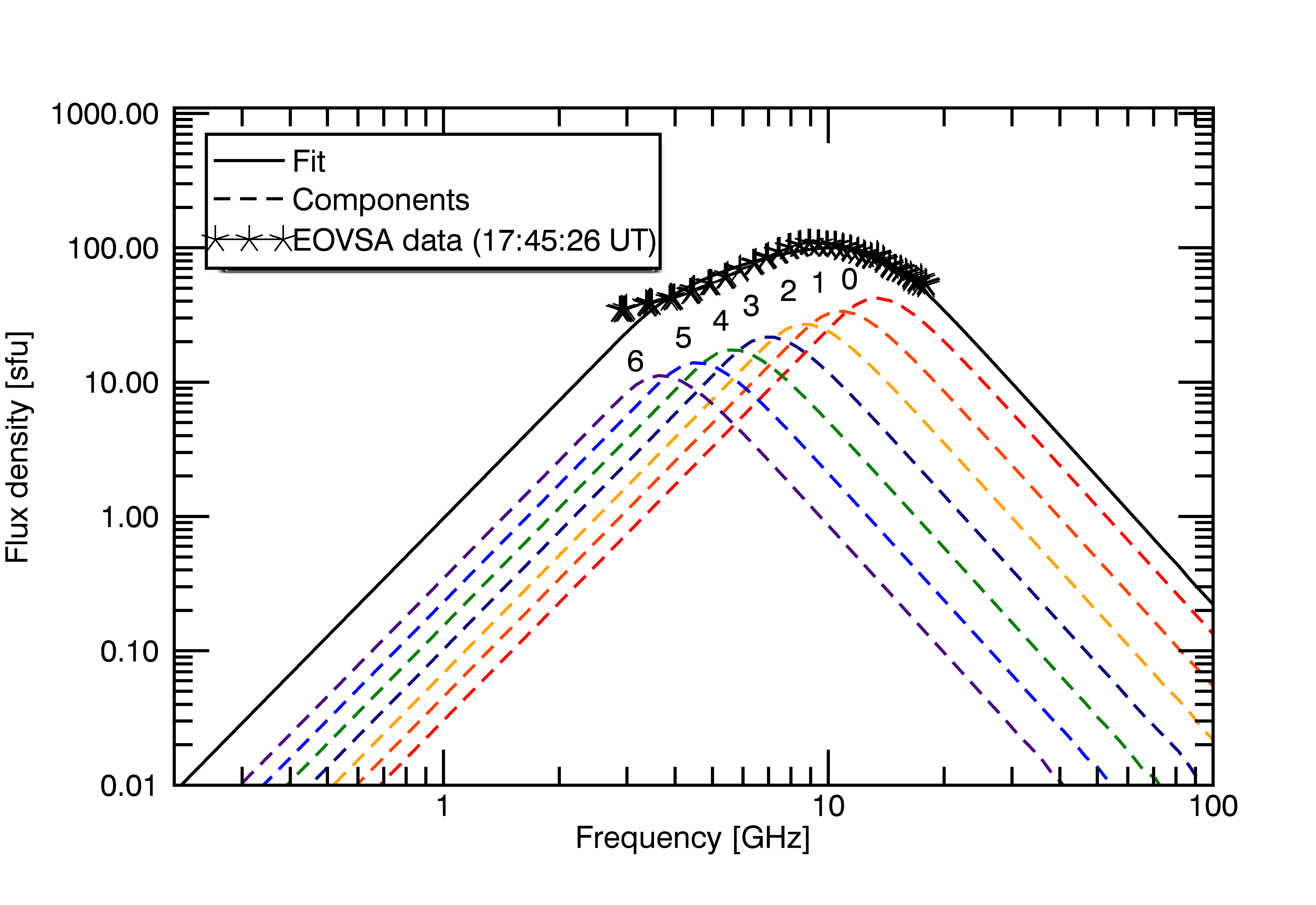}
\caption {Inhomogeneous model applied to the observed flux density spectrum of the 2015 August 24, M1.0 flare at the time marked. The dashed curves of various colors are the simultaneous emission spectra derived from the seven components labeled 0 to 6. The plus symbols mark the observed spectrum, and the thick black line shows the overall fit.}
\label{fig:fig8}
\end{figure}

As a specific example, in Figure~\ref{fig:fig8} we show the EOVSA total power spectrum of the 2015 August 24 event for the decay phase at 17:45:26~UT (marked with plus symbols). The dashed lines are the model spectra produced from each of the seven spectral components, and the black line shows the total contribution from all the emission of each component. These discrete components can be visualized as the emission from the source regions where the electrons have access, traveling from the main acceleration site during the flare. Individual peaks from these discrete sources are not expected to be observed in the total power flux density as the spectra are measured integrating over the area from the entire flaring region.

Along with the seven values of magnetic field $B_i$ (spaced between $1900$ and $380$ G), the parameters given in Table~\ref{table3} (except $\alpha$ and $\beta$) are varied according to their dependence on the spectral shape \citep{GaryH1989, StahliGH1989, GaryH2004}. The high frequency index $\alpha_h$ at the time of the spectrum gives electron power-law index $\delta$ $\approx$ $4.1$. Since $\alpha$ in Equation~\ref{eq:11} is the factor that controls the slope of the spectrum in the optically thick part, it is set separately from the other parameters. The remaining parameters, $B_0$ and $L_0$ are fixed with nominal values to obtain the peak frequency close to the observed one. Then, $A_0$ is set to match the total flux density, and $N_{NT}$ value is set arbitrarily, which is the spatial density of the non-thermal electrons. Having these multi-variant parameters, the model spectra that best matches the observed spectrum are distinguished having the least $\chi^2$ value generated from 
\[\chi^2 = \sum_{i=0}^{nf-1} \frac{[S(\nu_i)-S_m(\nu_i)]^2}{\sigma_i^2}, \]
where $S(\nu_i)$ and $S_m(\nu_i)$ are the observed and model fit flux densities with $\sigma_i$ being the uncertainty at each frequency $\nu_i$.

\begin{deluxetable*}{ccccccc}
\tablenum{3}
\tablecaption{Model parameters\label{table3}}
\tablewidth{0pt}
\tablehead{
\colhead{$\alpha$} & \colhead{$\beta$} & \colhead{$\delta$} & \colhead{$\theta^{\circ}$} & \colhead{Density $N_{NT}$ (${\rm cm}^{-3}$)} & \colhead{Thickness $L_6$ (arcsec)} & \colhead{Area $A_6$ (arcsec$^2$)}
}
\startdata
0.6 & 0.2 & 4.1 & 50 & ${3.8}\times{10^6}$ & $69$ & $284$
\enddata
\end{deluxetable*}

The area $A_6$ and the thickness $L_6$ in the table are for the lowest-frequency component 6, peaking at $\sim3.2$ GHz. The area and thickness of the components (0 to 6) range from 108 to 284 arcsec$^2$ and 50 to 69 arcsec, respectively. To resolve the smallest (11$\arcsec$) of these discrete sources requires a radio array with modest baseline lengths of order 1 km depending on frequency (0.38 km at 18 GHz). However, EOVSA imaging spectroscopy already provides a much higher resolution of 3.3$\arcsec$ at 18 GHz.

As the fluxes in this model are summed to match the flux density spectrum, when each component is combined, give an overall area of $\sim$1290 arcsec$^2$ (equivalent circular size of $\sim$40 arcsec). This size is although smaller than the size estimates given earlier but nevertheless serves to show that an inhomogeneous source can account for the shape of the spectrum. In addition, the corresponding emission volumes accounting for the LOS thickness, $V_j = A_jL_j$, for any component \textit{j}, are quite large and grow larger at low frequencies. 

These measurements indicate that for reproducing a flat spectrum, the emission either has to be comprised of multiple emission components simultaneously observed within the flare volume or has to be from a huge volume. The model shows that the flat spectrum can be the consequence of a significantly large source structure that is implausible to be homogeneous for such extended physical space over the active region and can only be inhomogeneous in nature.  

\section{Summary}
 We study the flare radio source morphology in the low-frequency emission using the flux density spectra of 12 bursts during 2015 with the excellent frequency and time resolution data available from the EOVSA interferometer. Having the optically thick spectral index as a proxy for MW source morphology, we illustrate the LF sources associated with the flat spectra by the following characteristics.

\begin{enumerate}
\item{A flat spectrum can be explained as the emission from spatially inhomogeneous, non-uniform physical parameters of a large source area and/or with simultaneous multiple emission components within. First, the relative visibility source area measurements have shown that the events with flat spectra have a source size greater than at least $\sim120\arcsec$ at low frequencies. Second, the observed flat spectrum can only be reproduced by the inhomogeneous model with discrete parameters on the source function. Finally, the area spectrum analysis indicates that the source size observed at a relatively low brightness temperature is still large in the case of a flat spectrum than the typically observed LF sources (that roughly follow $A \propto \nu^{-2}_{\rm GHz}$ \cite{Bastian1999}). Therefore, the MW sources at low-frequency can be large, extended, and complex in the spatial domain, whose existence suggests that the accelerated particles have access to a large region of space during the flare.}

\item{As an evolutionary trend, we observed that most spectra (for nine out of 12 events) exhibit a decrease in $\alpha_l$ significantly in the decay phase from the index value at the peak time.  This trend indicates that the inhomogeneity and the complexity of the emission volume increase as the flare process advances. As a piece of evidence for this from flare imaging, the event discussed in \cite{Gary2018} shows the sources during the decay phase multiply to a bigger size and discrete spatial characteristics.}

\item{Five of the flat events, in particular, have shown a more shallow and flatter spectrum. Their spectral index is much less than $1.0$ in at least any of the three phases of the bursts. In turn, we conclude that the occurrence of large and complex MW sources, i.e., indicated by the flat spectrum, can be seen in $42\%$ of the flares (5 out of 12 events). All of these flat spectral events are originated mainly from the active regions with a complex magnetic configuration of $\beta\gamma\delta$ (as in Table~\ref{table1}). We also find that a flat spectral event need not necessarily be a high-intensity flare with a huge flux density.} \end{enumerate}

In summary, focusing mainly on the low-frequency emission and flat spectral cases, this study has given the means to understand the characteristics of the seldom examined LF MW flare sources relative to the usually observed high frequency optically thin sources. The large volumes of these sources can involve the large-scale coronal loops filled with particles that get injected and escaped as the seeding particles for solar energetic particle (SEP) events.

A better understanding of these large LF sources and their role during a flare, their magnetic field structure, and their spatial relationship to more commonly observed components of solar flares can be achievable with adequate imaging data now becoming available. This work highlights the importance of focusing on the LF optically thick microwave emission in future studies.
\hfill \break
\\
\textbf{Acknowledgements} We thank Jeongwoo Lee for providing IDL code of the inhomogeneous model and giving his valuable suggestions on the model. SS thanks Kim Tolbert and late Richard Schwartz for their helpful discussions in fixing the issues with RHESSI imaging data of one of the events. This work was supported by NSF grant AST-1910354 and NASA grants 80NSSC18K1128 and 80NSSC19K0068 to New Jersey Institute of Technology.   

\bibliography{paper_bib}
\bibliographystyle{aasjournal}
\end{document}